\documentclass[11pt]{article}
\usepackage{amssymb,amsmath,epsfig,cite,graphicx,setspace}
\usepackage{dcolumn}
\usepackage{bm}
\usepackage{amsfonts} 
\usepackage{amssymb} 

\begin{document}

\title{\bf Roles of Modified Chaplygin-Jacobi and Chaplygin-Abel Gases in FRW Universe}

\author{Ujjal Debnath
\thanks{ujjaldebnath@gmail.com}\\
Department of Mathematics, Indian Institute of Engineering\\
Science and Technology, Shibpur, Howrah-711 103, India.\\}

\maketitle

\begin{abstract}
{\large We have considered flat Friedmann-Robertson-Walker (FRW)
model of the universe and reviewed the modified Chaplygin gas as
the fluid source. Associated with the scalar field model, we have
determined the Hubble parameter as a generating function in terms
of the scalar field. Instead of hyperbolic function, we have taken
Jacobi elliptic function and Abel function in the generating
function and obtained modified Chaplygin-Jacobi gas (MCJG) and
modified Chaplygin-Abel gas (MCAG) equation of states,
respectively. Next, we have assumed that the universe filled in
dark matter, radiation, and dark energy. The sources of dark
energy candidates are assumed as MCJG and MCAG. We have
constrained the model parameters by recent observational data
analysis. Using $\chi^{2}$ minimum test (maximum likelihood
estimation), we have determined the best fit values of the model
parameters by OHD+CMB+BAO+SNIa joint data analysis. To examine the
viability of the MCJG and MCAG models, we have determined the
values of the deviations of information criteria like
$\triangle$AIC, $\triangle$BIC and $\triangle$DIC. The evolutions
of cosmological and cosmographical parameters (like equation of
state, deceleration, jerk, snap, lerk, statefinder, Om diagnostic)
have been studied for our best fit values of model parameters. To
check the classical stability of the models, we have examined the
values of square speed of sound $v_{s}^{2}$ in the interval
$(0,1)$ for expansion of the universe.
}
\end{abstract}

\newpage

\sloppy \tableofcontents

\section{\normalsize\bf{Introduction}}

From Supernova type Ia \cite{Perlmutter,Perlmutter1,Riess1}
observation, it has well-established that our universe is
undergoing acceleration phase. This is also confirmed by other
observations like Baryon Acoustic Oscillation (BAO) \cite{EU,EU1},
Cosmic Microwave Background (CMB) \cite{Briddle}, WMAP
\cite{Sper2,Kom} and Planck \cite{Ade} observations. There is some
unknown matter which possesses sufficient negative pressure
(repulsive force), which derives the universe to the acceleration,
dubbed as dark energy (DE). According to the most simple and
successful candidate of dark energy is the cosmological constant
$\Lambda$ (or vacuum energy) and its density consist of $\sim$
70\% of total energy density in the universe
\cite{Pee,Padm,Cope,Cald0}. The concordance $\Lambda$ cosmology is
consistent with the cosmological observations, but it cannot solve
the fine-tuning, and the cosmic coincidence problems
\cite{Padm,Cope,Wein}. So other candidates of dark energy obey
time-evolving energy density have been proposed to solve the above
cosmological problems \cite{Arme00,Cald00,Eli,Cald}. The
quintessence (or scalar field)
\cite{Cald,Peebles,Tsu,Bam,Chen,Smer,Rat,Wet} is another most
accepted candidate of DE. There are other scalar field models
which are also the candidates of DE models
\cite{Arme,Sen,Feng,Guo,gas,Gum,Mart,Wei}. On the other hand,
modified theories of gravity may be the alternatives to the DE to
derive the acceleration of the universe
\cite{Dvali,Jacob,Ab,Noj,Hora,Barrow,Meng1,Tsu1}.\\

Another alternative to DE is an exotic type of fluid - the
so-called pure Chaplygin gas. The pure Chaplygin gas derives to
the present-day acceleration of the universe, which can be treated
as a unification of dark matter and DE \cite{Kam1,Gor,Bil}. Then
it has been generalized to the generalized Chaplygin gas, which
also unifies dark matter and DE \cite{Ben1,Mak,Gor1}. Further, it
has been modified to the modified Chaplygin gas (MCG), which
propagates between radiation and $\Lambda$CDM stages
\cite{Bena,DebAC}. Other extensions of Chaplygin gas are variable
Chaplygin gas \cite{Guo1}, variable modified Chaplygin gas
\cite{Debnn}, viscous Chaplygin gas \cite{Zhai} etc., which are
also the candidates of DE models that unify dark matter and DE.
Due the Supernova, CMBR, WMAP observations, the parameters of the
generalized Chaplygin gas have been constrained
\cite{Mak,Ben2,Ben3,Amen}. Also, the MCG parameters have been
constrained by CMBR, WMAP data \cite{LiuL,Lu}. On the other hand,
the MCG parameters have been constrained in modified gravity
theories
\cite{Chak,P1,D2,R1,D1,UD000}.\\

Villanueva \cite{Vill} has obtained generalized Chaplygin-Jacobi
gas through Hubble parameter associated to the generalized
Chaplygin scalar field and using Jacobi's elliptic function
\cite{Vill1}. The relevant inflationary quantities have also been
obtained \cite{Cad}. Motivated by this work, we obtain modified
Chaplygin-Jacobi gas and modified Chaplygin-Abel gas equation of
states. Then we study the observational data analysis and
constrains the model parameters. We check the compatibility of the
models with observational data employing AIC, BIC, and DIC model
selection tools. Using the best-fitting values of the models' free
parameters, we study the evolutions of cosmological and
cosmographical parameters and check the classical stability of the
models. The paper is structured as follows. In section 2, we
briefly write the basic equations for MCG. In sections 3 and 4, we
obtain modified Chaplygin-Jacobi gas and modified Chaplygin-Abel
gas equation of states through the Hubble parameter. In section 5,
we write the data analysis tools in our constructed models. Then
we study the OHD+CMB+BAO+SNIa joint data analysis and the
information criteria for our models. Section 4 deals with the
evolutions of cosmological and cosmographical parameters using
best-fitting values of the model parameters. Finally, in section
7, we discuss the results of the whole work.\\

\section{\normalsize\bf{Review of Modified Chaplygin Gas in FRW Model}}

The line element for homogeneous and isotropic universe in
spatially flat Friedmann-Robertson-Walker (FRW) model is (choosing
speed of light $c=1$)
\begin{equation}
ds^{2}=-dt^{2}+a^{2}(t)\left[dr^{2}+r^{2}(d\theta^{2}+sin^{2}\theta
d\phi^{2})\right]
\end{equation}
where $a(t)$ is a scale factor depends on time $t$. The Einstein's
field equations are given by
\begin{equation}\label{F1}
3H^{2}=8\pi G \rho
\end{equation}
and
\begin{equation}\label{F2}
\dot{H}=-4\pi G(\rho+p)
\end{equation}
where $\rho$ and $p$ are the energy density and pressure
respectively, $H=\dot{a}/a$ denotes the time dependent Hubble
parameter and $G$ is the Newtonian gravitational constant. The
energy conservation equation is given by
\begin{equation}\label{CE}
\dot{\rho}+3H(\rho+p)=0
\end{equation}
We consider the fluid source is modified Chaplygin gas (MCG),
which obeys the equation of state \cite{Bena,DebAC}
\begin{equation}\label{MCG}
p=A\rho-\frac{B}{\rho^{\alpha}}
\end{equation}
where $A$, $B(>0)$ and $\alpha\in(0,1)$ are constants. We have the
solution of $\rho$ as \cite{Bena,DebAC}
\begin{equation}\label{rho1}
\rho=\left[\frac{B}{1+A}+\frac{C}{a^{3(1+A)(1+\alpha)}}
\right]^{\frac{1}{1+\alpha}}
\end{equation}
where $C$ is an integration constant.\\

From field-theoretic point of view, we may describe this MCG model
by introducing scalar field $\phi$ and corresponding
self-interacting potential $V(\phi)$, which have the effective
Lagrangian
\begin{equation}
{\cal L_{\phi}}=\frac{1}{2}\dot{\phi}^{2}-V(\phi)
\end{equation}
The analogous energy density $\rho_{\phi}$ and pressure $p_{\phi}$
for the scalar field are the following \cite{DebAC}:
\begin{equation}\label{rphi}
\rho_{\phi}=\frac{1}{2}\dot{\phi}^{2}+V(\phi)=\rho
\end{equation}
and
\begin{equation}\label{pphi}
p_{\phi}=\frac{1}{2}\dot{\phi}^{2}-V(\phi)=A\rho-\frac{B}{\rho^{\alpha}}
\end{equation}
From these equations with the help of field equation (\ref{F1}),
we get the solution \cite{DebAC}:
\begin{equation}\label{phi1}
\phi=\phi_{c}+\frac{1}{\sqrt{6\pi
G(1+A)}(1+\alpha)}~Sinh^{-1}\left\{\sqrt{\frac{C(1+A)}{B}}\frac{1}{a^{\frac{3}{2}(1+\alpha)(1+A)}}
\right\}
\end{equation}
where $\phi_{c}$ is constant. Using equations (\ref{F1}),
(\ref{rho1}) and (\ref{phi1}), we obtain the generating function
for MCG
\begin{equation}\label{H}
H(\phi)=H_{c}~Sech^{-\frac{1}{1+\alpha}}(\Phi)
\end{equation}
where $\Phi=\sqrt{6\pi G(1+A)}(1+\alpha)(\phi-\phi_{c})$ and
$H_{c}=\sqrt{\frac{8\pi
G}{3}}\left(\frac{B}{1+A}\right)^{\frac{1}{2(1+\alpha)}}$ is the
value of the Hubble parameter when $\phi=\phi_{c}$. Now from
equation (\ref{F2}), we get
\begin{equation}\label{prho}
p=-\rho+\frac{1}{(4\pi G)^{2}}\left(\frac{dH}{d\phi}\right)^{2}
\end{equation}

\section{\normalsize\bf{Modified Chaplygin-Jacobi Gas}}

We know that hyperbolic function is a particular case of elliptic
functions and Jacobi elliptic functions are a set of basic
elliptic functions. In 2015, Villanueva \cite{Vill} has replaced
the cosine hyperbolic function in the Hubble parameter (obtained
by generalized Chaplygin gas) by Jacobi's elliptic function and
derived the equation of state, termed as generalized
Chaplygin-Jacobi gas. So replacing the hyperbolic function in the
generating function (\ref{H}) by the Jacobi elliptic cosine
function $cn(\Phi)$, the generating function can be written as
\begin{equation}\label{HJ}
H(\phi)=H_{c}~cn^{-\frac{1}{1+\alpha}}(\Phi)
\end{equation}
where $cn(\Phi)\equiv cn(\Phi,k)$ and $k\in[0,1]$ is the elliptic
modulus. For $k=1$, the above equation (\ref{HJ}) reduces to the
equation (\ref{H}). Using the relations
$cn^{2}(\Phi)+sn^{2}(\Phi)=1,~cn^{2}(\Phi)+(1-k)sn^{2}(\Phi)=dn^{2}(\Phi),~cn'(\Phi)=-sn(\Phi)~dn(\Phi)$
(the notations are defined in \cite{Vill}) and taking the
derivative of (\ref{HJ}), we obtain
\begin{equation}\label{HJ2}
\left(\frac{dH}{d\phi}\right)^{2}=6\pi
G(1+A)H_{c}^{2}cn^{-\frac{2(2+\alpha)}{1+\alpha}}(\Phi)(1-cn^{2}(\Phi))((1-k)+k~cn^{2}(\Phi))
\end{equation}
Also using equations (\ref{rphi}), (\ref{pphi}), (\ref{prho}),
(\ref{HJ}) and (\ref{HJ2}) we can obtain the potential function
\begin{equation}
V(\phi)=\frac{1}{2}\left(\frac{B}{1+A}\right)^{\frac{1}{2(1+\alpha)}}~cn^{-\frac{2}{2+\alpha}}(\Phi)
[2+(1-cn^{-2}(\Phi))((1-k)+k~cn^{2}(\Phi))]
\end{equation}
Using equations (\ref{F1}), (\ref{prho}), (\ref{HJ}) and
(\ref{HJ2}), we finally obtain the relation between pressure and
density as in the form
\begin{equation}\label{MCJG}
p=[(2k-1)(1+A)-1]\rho-\frac{kB}{\rho^{\alpha}}+\frac{(1-k)(1+A)^{2}}{B}~\rho^{2+\alpha}
\end{equation}
which may be called {\it ``Modified Chaplygin-Jacobi Gas"} (MCJG)
equation of state. For $A=0$, the MCJG equation of state can be
reduced to the generalized Chaplygin-Jacobi gas equation of state
\cite{Vill}. Putting the expression of $p$ from equation
(\ref{MCJG}) into (\ref{CE}), we obtain
\begin{equation}
\rho^{1+\alpha}=\frac{B}{1+A}\left[\frac{a^{3(1+\alpha)(1+A)}+kD}{a^{3(1+\alpha)(1+A)}-(1-k)D}
\right]
\end{equation}
where $D$ is a positive constant. For large value of scale factor
$a(t)$, we have $ \rho\simeq
\left(\frac{B}{1+A}\right)^{\frac{1}{1+\alpha}}~\text{and}~p\simeq
-\left(\frac{B}{1+A}\right)^{\frac{1}{1+\alpha}}=-\rho $ which
corresponds to the $\Lambda$CDM model with the cosmological
constant $=\left(\frac{B}{1+A}\right)^{\frac{1}{1+\alpha}}$. But
on the other hand, in the phase of the universe where
$x=a^{3(1+\alpha)(1+A)}-(1-k)D \simeq 0$, we have $ \rho\simeq
\left(\frac{B}{1+A}~\frac{D}{x}\right)^{\frac{1}{1+\alpha}} $.
This corresponds to a phase of the universe which follows
polytropic equation of state
$p=\frac{(1-k)(1+A)}{B}\rho^{\alpha+2}$. The above expression of
$\rho$ can be written in the following form
\begin{equation}\label{rhoMCJG}
\rho^{1+\alpha}=\frac{B}{1+A}\left[\frac{A_{s}+(1-A_{s})(1+z)^{3(1+\alpha)(1+A)}}
{kA_{s}-(1-k)(1-A_{s})(1+z)^{3(1+\alpha)(1+A)}} \right]
\end{equation}
where $A_{s}=\frac{1}{1+kD}$ satisfying $1-k<A_{s}<1$. So the
present value of the energy density is
$\rho_{CJ0}^{1+\alpha}=\frac{B}{(1+A)[kA_{s}-(1-k)(1-A_{s})]}$.
The equation of state (EoS) parameter is obtained in the form
\begin{eqnarray}
W=\frac{p}{\rho}=[(2k-1)(1+A)-1]-k(1+A)\left[\frac{kA_{s}-(1-k)(1-A_{s})(1+z)^{3(1+\alpha)(1+A)}}{A_{s}+(1-A_{s})(1+z)^{3(1+\alpha)(1+A)}}\right]
\nonumber
\\
+(1-k)(1+A)\left[\frac{A_{s}+(1-A_{s})(1+z)^{3(1+\alpha)(1+A)}}
{kA_{s}-(1-k)(1-A_{s})(1+z)^{3(1+\alpha)(1+A)}} \right]
\end{eqnarray}

\section{\normalsize\bf{Modified Chaplygin-Abel Gas}}

Abel elliptic functions are a special kind of elliptic function.
If we replace the hyperbolic function in the generating function
(\ref{H}) by the Abel elliptic function $F(\Phi)$, the generating
function (\ref{H}) in terms of Abel elliptic function can be
written as
\begin{equation}\label{HA}
H(\phi)=H_{c}~F^{-\frac{1}{1+\alpha}}(\Phi)
\end{equation}
Here $F(\Phi)=\sqrt{1+e^{2}\varphi^{2}(\Phi)}$ where
$\varphi(\Phi)\equiv \varphi(\Phi,c,e)$ is also Abel elliptic
function (here $c,e\in \mathbb{R}$). Using the relation
$\varphi'(\Phi)=\sqrt{(1-c^{2}\varphi^{2}(\Phi))(1+e^{2}\varphi^{2}(\Phi))}$
and taking the derivative of (\ref{HA}), we obtain
\begin{equation}\label{HA2}
\left(\frac{dH}{d\phi}\right)^{2}=6\pi
G(1+A)H_{c}^{2}F^{-\frac{2(2+\alpha)}{1+\alpha}}(\Phi)(F^{2}(\Phi)-1)((c^{2}+e^{2})-c^{2}F^{2}(\Phi))
\end{equation}
Also using equations (\ref{rphi}), (\ref{pphi}), (\ref{prho}),
(\ref{HA}) and (\ref{HA2}) we can obtain the potential function
\begin{equation}
V(\phi)=\frac{1}{2}\left(\frac{B}{1+A}\right)^{\frac{1}{2(1+\alpha)}}~F^{-\frac{2}{2+\alpha}}(\Phi)
[2+(F^{-2}(\Phi)-1)((c^{2}+e^{2})-c^{2}F^{2}(\Phi))]
\end{equation}
Using equations (\ref{F1}), (\ref{prho}), (\ref{HA}) and
(\ref{HA2}), we finally obtain the relation between pressure and
density as in the form
\begin{equation}\label{MCAG}
p=[(e^{2}+2c^{2})(1+A)-1]\rho-\frac{c^{2}B}{\rho^{\alpha}}-\frac{(e^{2}+c^{2})(1+A)^{2}}{B}~\rho^{2+\alpha}
\end{equation}
which may be called {\it ``Modified Chaplygin-Abel Gas"} (MCAG)
equation of state. For $A=0$, the MCAG equation of state may be
called the generalized Chaplygin-Abel gas equation of state.
Putting the expression of $p$ from equation (\ref{MCAG}) into
(\ref{CE}), we obtain
\begin{equation}
\rho^{1+\alpha}=\frac{B}{1+A}\left[\frac{a^{3e^{2}(1+\alpha)(1+A)}+c^{2}K}{a^{3e^{2}(1+\alpha)(1+A)}+(e^{2}+c^{2})K}
\right]
\end{equation}
where $K$ is a positive constant. For large value of scale factor
$a(t)$, we obtain $p\simeq -\rho$ which corresponds to the
$\Lambda$CDM model with the cosmological constant
$=\left(\frac{B}{1+A}\right)^{\frac{1}{1+\alpha}}$. For small
value of scale factor $a(t)$, we also obtain $p\simeq -\rho$,
which corresponds to the inflationary stage of the universe. So
the MCAG propagates between inflation and $\Lambda$CDM stages. The
above expression of $\rho$ can be written in the form
\begin{equation}\label{rhoMCAG}
\rho^{1+\alpha}=\frac{c^{2}B}{1+A}\left[\frac{B_{s}+(1-B_{s})(1+z)^{3e^{2}(1+\alpha)(1+A)}}
{c^{2}B_{s}+(e^{2}+c^{2})(1-B_{s})(1+z)^{3e^{2}(1+\alpha)(1+A)}}
\right]
\end{equation}
where $B_{s}=\frac{1}{1+c^{2}K}$ satisfying $0<B_{s}<1$. So the
present value of the energy density is
$\rho_{CA0}^{1+\alpha}=\frac{c^{2}B}{(1+A)[c^{2}B_{s}+(e^{2}+c^{2})(1-B_{s})]}$.
The equation of state (EoS) parameter is obtained in the form
\begin{eqnarray}
W=\frac{p}{\rho}=[(e^{2}+2c^{2})(1+A)-1]-(1+A)\left[\frac{c^{2}B_{s}+(e^{2}+c^{2})(1-B_{s})(1+z)^{3e^{2}(1+\alpha)(1+A)}}
{B_{s}+(1-B_{s})(1+z)^{3e^{2}(1+\alpha)(1+A)}} \right] \nonumber
\\
-c^{2}(e^{2}+c^{2})(1+A)\left[\frac{B_{s}+(1-B_{s})(1+z)^{3e^{2}(1+\alpha)(1+A)}}
{c^{2}B_{s}+(e^{2}+c^{2})(1-B_{s})(1+z)^{3e^{2}(1+\alpha)(1+A)}}
\right]
\end{eqnarray}

\section{Observational Data Analysis}

Now, we assume that the universe is composed of radiation, dark
matter (DM), and dark energy (DE). So in the Einstein field
equations (\ref{F1}), (\ref{F2}) and in the conservation equation
(\ref{CE}), we can consider $\rho=\rho_{r}+\rho_{m}+\rho_{d}$ and
$p=p_{r}+p_{m}+p_{d}$. Here suffices $r,~m,~d$ denote the
radiation, dark matter, and dark energy, respectively. If we
assume that there is no interaction between radiation, DM, and DE,
then their conservation equations will be
\begin{equation}\label{rad}
\dot{\rho}_{r}+3H(\rho_{r}+p_{r})=0,
\end{equation}
\begin{equation}\label{DM}
\dot{\rho}_{m}+3H(\rho_{m}+p_{m})=0
\end{equation}
and
\begin{equation}\label{DE}
\dot{\rho}_{d}+3H(\rho_{d}+p_{d})=0
\end{equation}
For radiation, the EoS is $p_{r}=\frac{1}{3}\rho_{r}$, so from
equation (\ref{rad}) we get the solution
$\rho_{r}=\rho_{r0}(1+z)^{4}$. Also since the DM has negligible
pressure (i.e., $p_{m}\approx 0$), so from equation (\ref{DM}) we
obtain the solution $\rho_{m}=\rho_{m0}(1+z)^{3}$. Here
$\rho_{r0}$ and $\rho_{m0}$ represent the present values of
radiation density and DM density, respectively. Next we assume that
the DE of the universe is in the form of MCJG or MCAG.\\

$\bullet{}$ {\bf MCJG :} If the DE of the universe obeys the MCJG
equation of state (\ref{MCJG}), then using the equation
(\ref{rhoMCJG}), the equation (\ref{F1}) reduces to the following
equation for $E(z)$ as
\begin{eqnarray}
E^{2}(z)=\Omega_{r0}(1+z)^{4}+\Omega_{m0}(1+z)^{3}
+\Omega_{CJ0}[kA_{s}-(1-k)(1-A_{s})]^{\frac{1}{1+\alpha}}
\nonumber\\
\times\left[\frac{A_{s}+(1-A_{s})(1+z)^{3(1+\alpha)(1+A)}}
{kA_{s}-(1-k)(1-A_{s})(1+z)^{3(1+\alpha)(1+A)}}
\right]^{\frac{1}{1+\alpha}}
\end{eqnarray}
where we have defined the normalized Hubble parameter
$E(z)=\frac{H(z)}{H_{0}}$ and the dimensionless density parameters
$\Omega_{r0}=\frac{8\pi G\rho_{r0}}{3H_0^2}$,
$\Omega_{m0}=\frac{8\pi G\rho_{m0}}{3H_0^2}$ and
$\Omega_{CJ0}=\frac{8\pi G\rho_{_{CJ0}}}{3H_0^2}$. Putting $z=0$
in the above equation, we obtain
$\Omega_{r0}+\Omega_{m0}+\Omega_{CJ0}=1$.\\

$\bullet{}$ {\bf MCAG :} If the DE of the universe obeys the MCAG
equation of state (\ref{MCAG}), then using the equation
(\ref{rhoMCAG}), the equation (\ref{F1}) reduces to the following
equation for $E(z)$ as
\begin{eqnarray}
E^{2}(z)=\Omega_{r0}(1+z)^{4}+\Omega_{m0}(1+z)^{3}
+\Omega_{CA0}[c^{2}B_{s}+(e^{2}+c^{2})(1-B_{s})]^{\frac{1}{1+\alpha}}
\nonumber\\
\times\left[\frac{B_{s}+(1-B_{s})(1+z)^{3e^{2}(1+\alpha)(1+A)}}
{c^{2}B_{s}+(e^{2}+c^{2})(1-B_{s})(1+z)^{3e^{2}(1+\alpha)(1+A)}}
\right]^{\frac{1}{1+\alpha}}
\end{eqnarray}
where the dimensionless density parameter $\Omega_{CA0}=\frac{8\pi
G\rho_{_{CA0}}}{3H_0^2}$. Putting $z=0$ in the above equation, we
obtain $\Omega_{r0}+\Omega_{m0}+\Omega_{CA0}=1$.\\

In this section, we'll study the observational data analysis tools
for fitting the theoretical MCJG and MCAG models in flat FRW
universe. To determine the best fit values of the unknown
parameters of the models, we use the observed Hubble data (OHD),
cosmic microwave background (CMB) data, baryonic acoustic
oscillations (BAO) data and type Ia supernova (SNIa) data survey.
For this purpose, we use minimum values of $\chi^{2}$ with maximum
likelihood analysis using different confidence levels (like
$1\sigma,~ 2\sigma$ and $3\sigma$) due to OHD+CMB+BAO+SNIa
observational data.

\subsection{OHD}

Due to latest observed Hubble data (OHD) sets compilation
\cite{YuR,More,Maga} with 57 data points in the region
$0.07\lesssim z \lesssim 2.42$ \cite{Sha}, the $\chi^{2}$ value
(as a sum of standard normal distribution) can be written as
\begin{equation}
{\chi}^{2}_{OHD}=\sum\frac{(H(z)-H_{obs}(z))^{2}}{\sigma^{2}_{obs}(z)}
\end{equation}
where $H(z)$ and $H_{obs}(z)$ respectively represent the
theoretical and observed values of the Hubble parameter. Also
$\sigma_{obs}(z)$ refers to the standard error in the observed
value of $H$. Due to Planck 2015 results \cite{Ade}, we take a
prior for the present value of the Hubble constant
$H_{0}=67.8$km/s/MPc.

\subsection{CMB Data}

To constraints the dark energy, we use the Cosmic Microwave
Background (CMB) anisotropy spectrum at the last scattering
surface at the redshift $z\simeq 1100$. So we use the Planck
distance priors from the Planck 2015 data sets \cite{Ade}. The
proper angular diameter distance $D_{A}(z)$, CMB shift parameter
${\cal R}$, comoving sound horizon size $r_{s}(z)$ and acoustic
scale parameter $\ell_{A}$ are respectively given by \cite{XZ,WD}
\begin{equation}
D_{A}(z)=\frac{1}{H_{0}(1+z)}\int_{0}^{z}\frac{dz'}{E(z')}~,
\end{equation}
\begin{equation}
{\cal R}=H_{0}\sqrt{\Omega_{m0}}~(1+z_{*})D_{A}(z_{*})~,
\end{equation}
\begin{equation}\label{rs}
r_{s}(z)=\frac{1}{H_{0}}\int_{0}^{x}\frac{dx'}{x'E(x')\sqrt{3(1+\bar{R}_{b}x')}}
\end{equation}
and
\begin{equation}
\ell_{A}=(1+z_{*})~\frac{\pi D_{A}(z_{*})}{r_{s}(z_{*})}
\end{equation}
where $\bar{R}_{b}x=0.75 \rho_{b}/\rho_{\gamma}$. The $\rho_{b}$
and $\rho_{\gamma}$ are respectively the baryon energy density and
photon energy density at the present epoch. Also, the baryon
density is $\omega_{b}=\Omega_{b0}h^{2}$ \cite{XZ,WHu}. The
$z_{*}$ is the redshift in the CMB power spectrum at the last
scattering surface, given by the fitting formula \cite{XZ,WHu}
\begin{equation}
z_{*}=1048\left[1+0.00124\left(\Omega_{b0}h^{2}\right)^{-0.738}\right]
\left[1+\frac{0.0783\left(\Omega_{b0}h^{2}\right)^{-0.238}}{1+39.5\left(\Omega_{b0}h^{2}\right)^{-0.76}}~
\left(\Omega_{m0}h^{2}\right)^{\frac{0.560}{1+21.1\left(\Omega_{b0}h^{2}\right)^{1.81}}}\right]
\end{equation}
The $\chi^{2}$ function for CMB measurement, is defined by
\cite{Ade,XZ}
\begin{equation}
\chi^{2}_{CMB}=\triangle S_{i}\left[Cov_{CMB}^{-1}(S_{i},S_{j})
\right]\triangle S_{j},~~\triangle
S_{i}=S_{i}^{theory}-S_{i}^{obs}
\end{equation}
where $S_{1}=\ell_{A}$, $S_{2}={\cal R}$, $S_{3}=\omega_{b}$ and
$Cov^{-1}$ is the inverse covariance matrix. The CMB data does not
provide the full Planck information, but it is an optimum way of
studying several dark energy models.

\subsection{BAO Data}

Eisenstein et al. \cite{EU,EU1} proposed the Baryon Acoustic
Oscillation (BAO) observational data analysis. BAO refers to the
imprint left by relativistic sound waves in the early universe,
providing an observable to the late-time large-scale structure.
BAO measures the structures in the universe from very large scales
at different times (i.e., redshifts). For BAO measurement, the
effective distance measure $D_{V}(z)$ is defined by \cite{XZ,WD}
\begin{equation}
D_{V}(z)=\left[(1+z)^{2}D_{A}^{2}(z)~\frac{z}{H(z)} \right]^{1/3}
\end{equation}
The comoving sound horizon size $r_{s}(z_{d})$ is given in
equation (\ref{rs}), where $z_{d}$ is the redshift at the drag
epoch given by \cite{XZ,EU}
\begin{equation}
z_{d}=\frac{1291\left(\Omega_{m0}h^{2}\right)^{0.251}}{1+0.659\left(\Omega_{m0}h^{2}\right)^{0.828}}~
\left[1+0.313\left(\Omega_{m0}h^{2}\right)^{-0.419}\left(\Omega_{b0}h^{2}\right)^{0.238\left(\Omega_{m0}h^{2}\right)^{0.223}}\left[1+0.607\left(\Omega_{m0}h^{2}\right)^{0.674}
\right] \right]
\end{equation}
For our data analysis, we use four data points: (i) for 6dF Galaxy
Survey \cite{Beu}, $r_{s}(z_{d})/D_{V}(0.106)=0.336\pm 0.015$;
(ii) for SDSS-DR7 \cite{Ros}, $D_{V}(0.15)=(664\pm
25Mpc)(r_{d}/r_{d,fid})$; (iii) for BOSS-DR11 \cite{And},
$D_{V}(0.32)=(1264\pm 25Mpc)(r_{d}/r_{d,fid})$ and (iv) for
BOSS-DR11 \cite{And}, $D_{V}(0.57)=(2056\pm
20Mpc)(r_{d}/r_{d,fid})$. Th $\chi^{2}$ function for BAO
measurement, is defined by \cite{XZ}
\begin{equation}
\chi^{2}_{BAO}=\triangle S_{i}\left[Cov_{BAO}^{-1}(S_{i},S_{j})
\right]\triangle S_{j},~~\triangle
S_{i}=S_{i}^{theory}-S_{i}^{obs}
\end{equation}

\subsection{SNIa Data}

The most important and key class of cosmological probes is the
standard candles, in which most known standard candles are
Supernovae Type Ia (SNIa). For SNIa observation, the luminosity
distance $d_{L}(z)$ and distance modulus $\mu(z)$ are given by
\cite{Riess1,Perlmutter1,Kowalaski,Sco}
\begin{equation}
d_{L}(z)=(1+z)\int_{0}^{z}\frac{dz'}{E(z')}
\end{equation}
and
\begin{equation}
\mu(z)=5\log_{10} \left[\frac{d_{L}(z)/H_{0}}{1~MPc}\right]+25
\end{equation}
From SDSS-II sample, SNLS sample and few high redshift samples
from HST (Hubble Space Telescope) with 740 data points \cite{Bet}
with redshift $z$ lies in the region $0.01 \le z \le 1.30$, the
$\chi^{2}$ function for JLA (joint light-curve analysis)
compilation can be written as \cite{Amanullah,Mamon1,Cai1,DB}
\begin{equation}
\chi^{2}_{SN}=A_{SN}-\frac{B_{SN}^{2}}{C_{SN}}
\end{equation}
where
\begin{equation}
A_{SN}=\sum_{i=1}^{740}\frac{[\mu_{th}(z_{i})-\mu_{obs}(z_{i})]^{2}}{\sigma^{2}(z_{i})}~,
~B_{SN}=\sum_{i=1}^{740}\frac{[\mu_{th}(z_{i})-\mu_{obs}(z_{i})]}{\sigma^{2}(z_{i})}~,
~C_{SN}=\sum_{i=1}^{740}\frac{1}{\sigma^{2}(z_{i})}
\end{equation}
The total $\chi^2$ function for OHD+CMB+BAO+SNIa joint data
analysis can be written as
\begin{eqnarray}
\chi^2=\chi^2_{OHD}+\chi^2_{CMB}+\chi^2_{BAO}+\chi^{2}_{SN}
\end{eqnarray}
Using the $\chi^2$ minimum test, we can determine the best fit
values of the parameters of MCJG and MCAG in the FRW universe.
Since the MCJG and MCAG models have a different number of
parameters, so it is very difficult to a fair comparison between
these DE models by directly comparing their $\chi^{2}$ values.
Obviously, a DE model involves more parameters that have a lower
value of $\chi^{2}$. So to discuss the fair model comparison,
we need to study the information criteria in the following subsection.\\

\subsection{Information Criteria}

Since the MCJG and MCAG have several parameters, so to discuss the
analysis, we need to know the information criteria (IC). The
Akaike Information Criteria (AIC) \cite{Ak}, Bayesian Information
Criteria (BIC) \cite{Sc} and Deviance Information Criterion (DIC)
\cite{Spie,Lidd} are more popular among ICs. Since AIC was derived
by an approximate minimization of Kullback-Leibler information, so
AIC is an asymptotically unbiased estimator of Kullback-Leibler
information. In the standard assumption of Gaussian errors, the
corresponding Gaussian estimator for the AIC can be written as
\cite{Ander,Bur,Lidd,Su} $\text{AIC}=-2\ln ({\cal
L}_{max})+2\kappa+\frac{2\kappa(\kappa+1)}{N-\kappa-1}$ where
${\cal L}_{max}$ is the maximum likelihood function, $\kappa$ is
the number of model parameters, and $N$ is the number of data
points used in the data fit. Since $N\gg 1$, so the above
expression reduces to the original AIC version: $\text{AIC}=-2\ln
({\cal L}_{max})+2\kappa$. Also, the Bayesian estimator for BIC
can be defined as \cite{Ander,Bur,Lidd} $\text{BIC}=-2\ln ({\cal
L}_{max})+\kappa\ln N$. We see that the BIC is similar to AIC with
a different second term. For the given set of models, the
deviations of the IC values are
$\triangle\text{AIC}=\text{AIC}_{model}-\text{AIC}_{min}=\triangle\chi^{2}_{min}+2\triangle\kappa$
and
$\triangle\text{BIC}=\text{BIC}_{model}-\text{BIC}_{min}=\triangle\chi^{2}_{min}+\triangle\kappa
\ln N$. For data analysis, the extensively favorable range of
$\triangle\text{AIC}$ is $(0,2)$. The less support range of
$\triangle\text{AIC}$ is $(4,7)$, but when
$\triangle\text{AIC}>10$, the model provides no support. For a
model, $\triangle\text{BIC}$ of 2 provides positive evidence
against the model with higher BIC while $\triangle\text{BIC}$ of 6
supports strong evidence. On the other hand, the concepts of both
Bayesian statistics and information theory, the DIC can be defined
as \cite{Spie,Lidd}
$\text{DIC}=2\overline{D(\theta)}-D(\bar{\theta})$ where
$D(\theta)=-2\ln ({\cal L}(\theta))+C$. Here $C$ is a standard
constant depending only on the data which will vanish from any
derived quantity, $D$ is the deviance of the likelihood function,
and bar denotes the mean value of the posterior distribution. For
deviation of the DIC value
$\triangle\text{DIC}=\text{DIC}_{model}-\text{DIC}_{min}$, the
lower value $(<2)$ provides strong support of the model, instead
of the higher value of $\triangle\text{DIC}$ (less support). It
should be noted that the AIC and BIC count all the involved
parameters of the model, but DIC counts only the number of
parameters of the model that contribute to the fit in an actual way.\\

\begin{figure}
~~~~~~~~~~
\includegraphics[height=2.5in, width=2.7in]{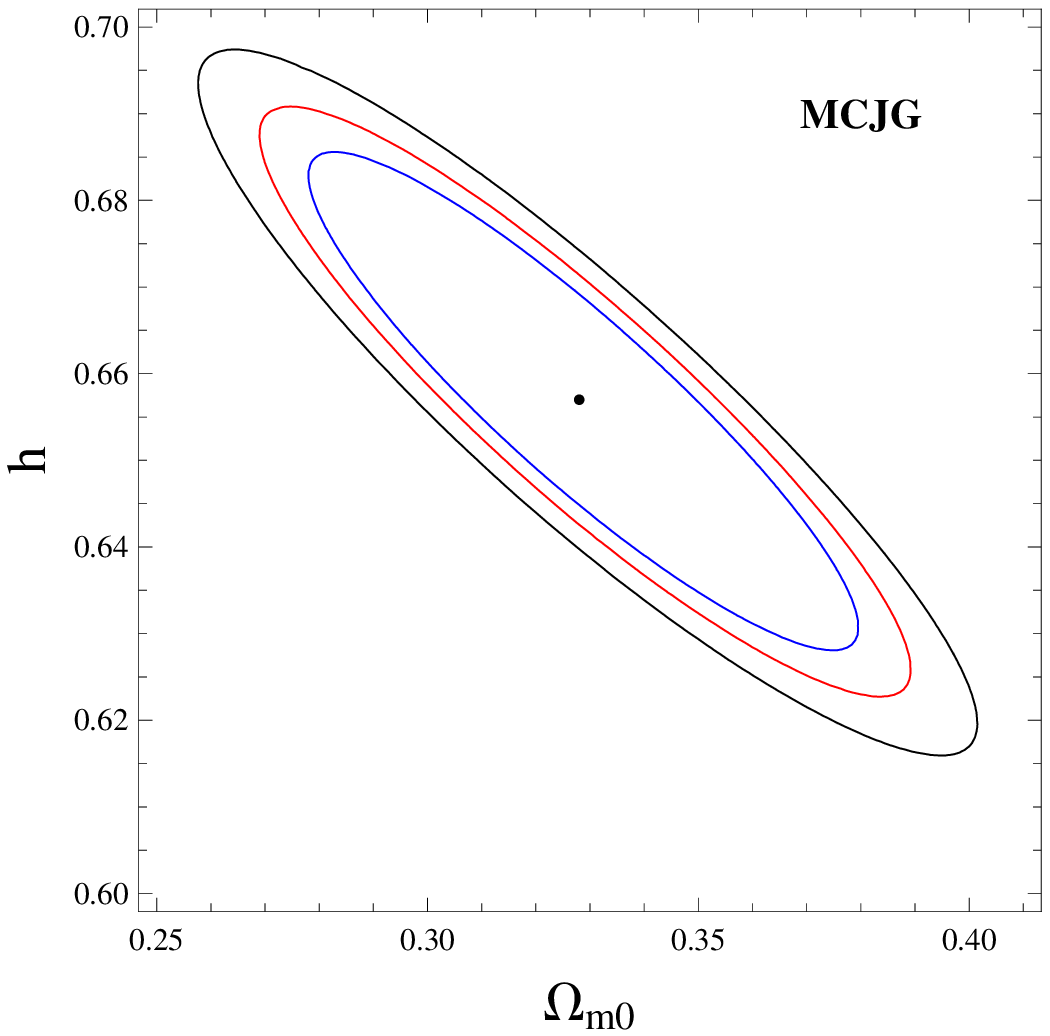}~~~~~~~
\includegraphics[height=2.5in, width=2.7in]{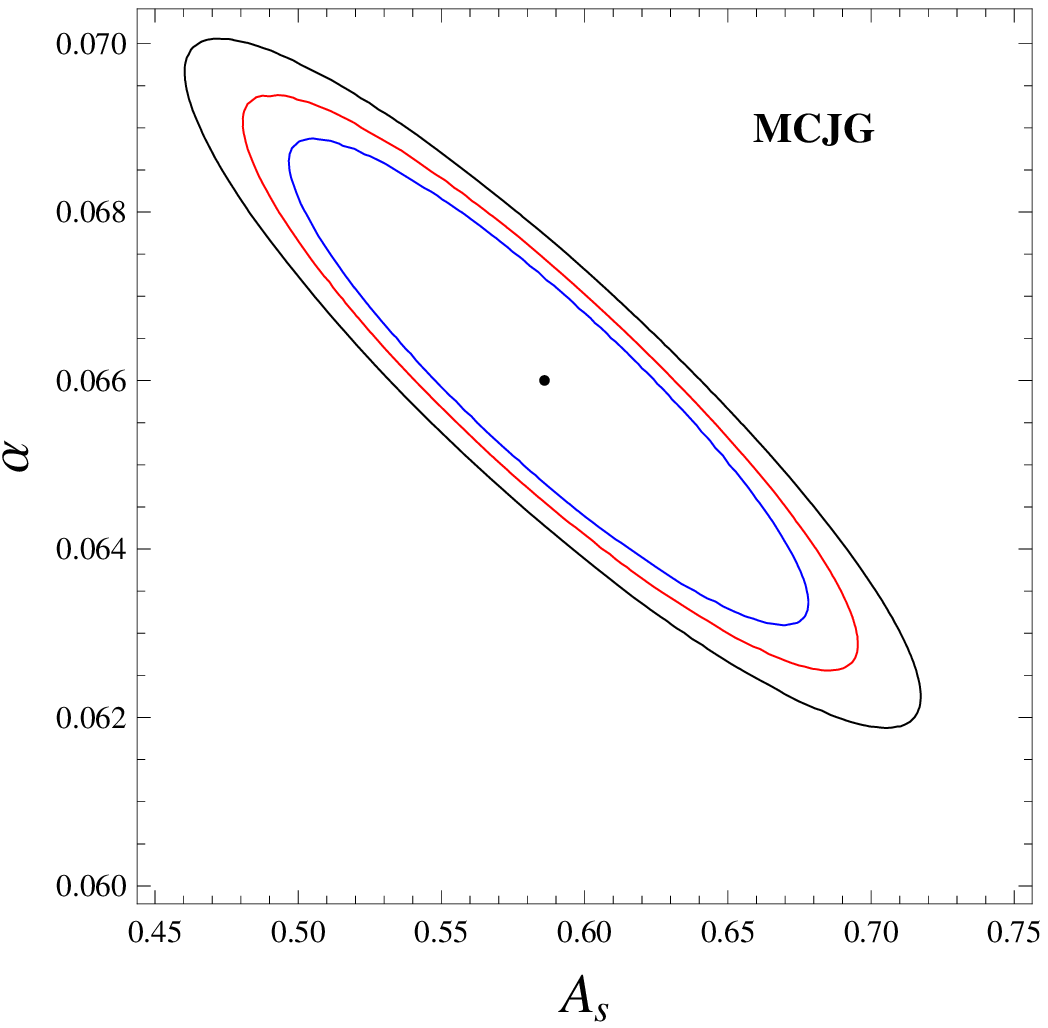}~~\\
\vspace{2mm}
~~~~~~~~~~~~~~~~~~~~~~~~~~~~~~~~~~~~Fig.1~~~~~~~~~~~~~~~~~~~~~~~~~~~~~~~~~~~~~~~~~~~~~~~~~~~~~~~~Fig.2
\\
\vspace{2mm} ~~~~~~~~~~
\includegraphics[height=2.5in, width=2.7in]{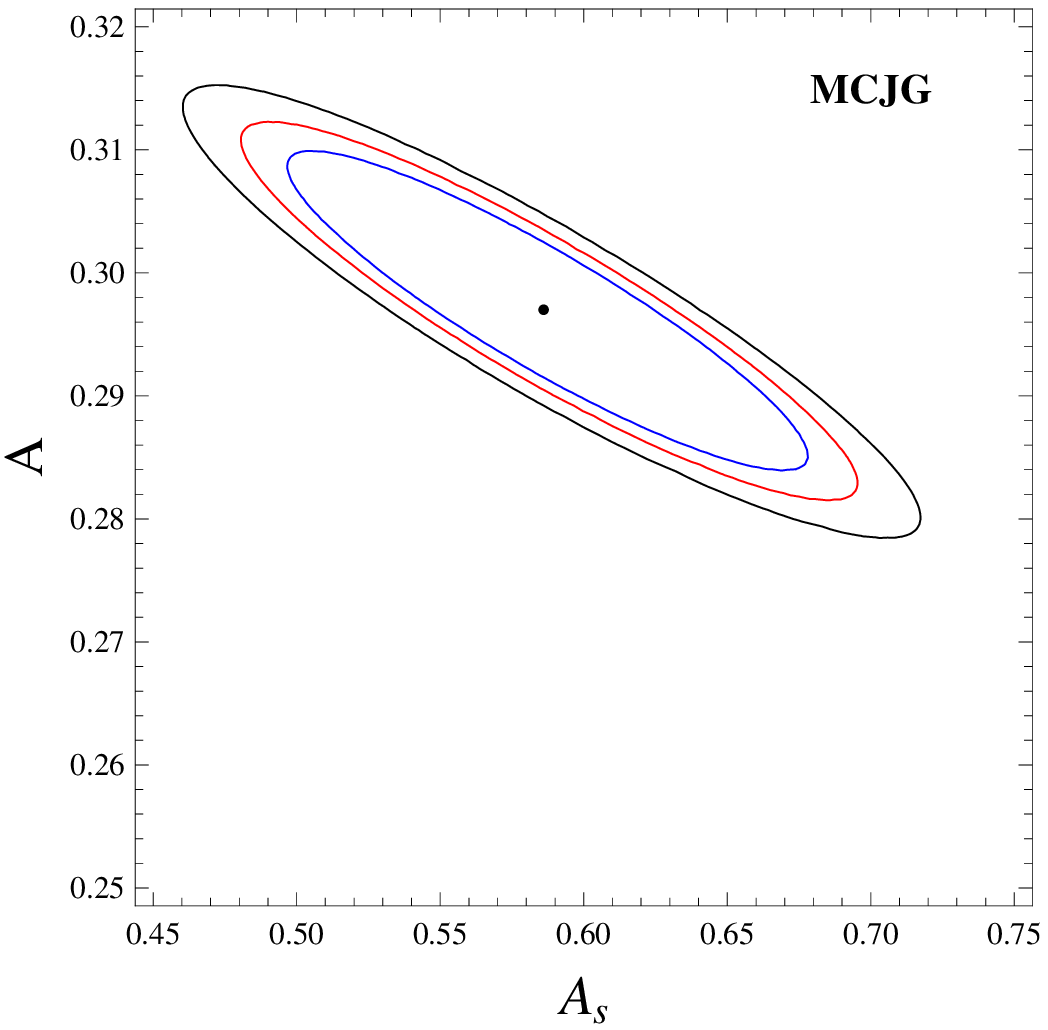}~~~~~~~
\includegraphics[height=2.5in, width=2.7in]{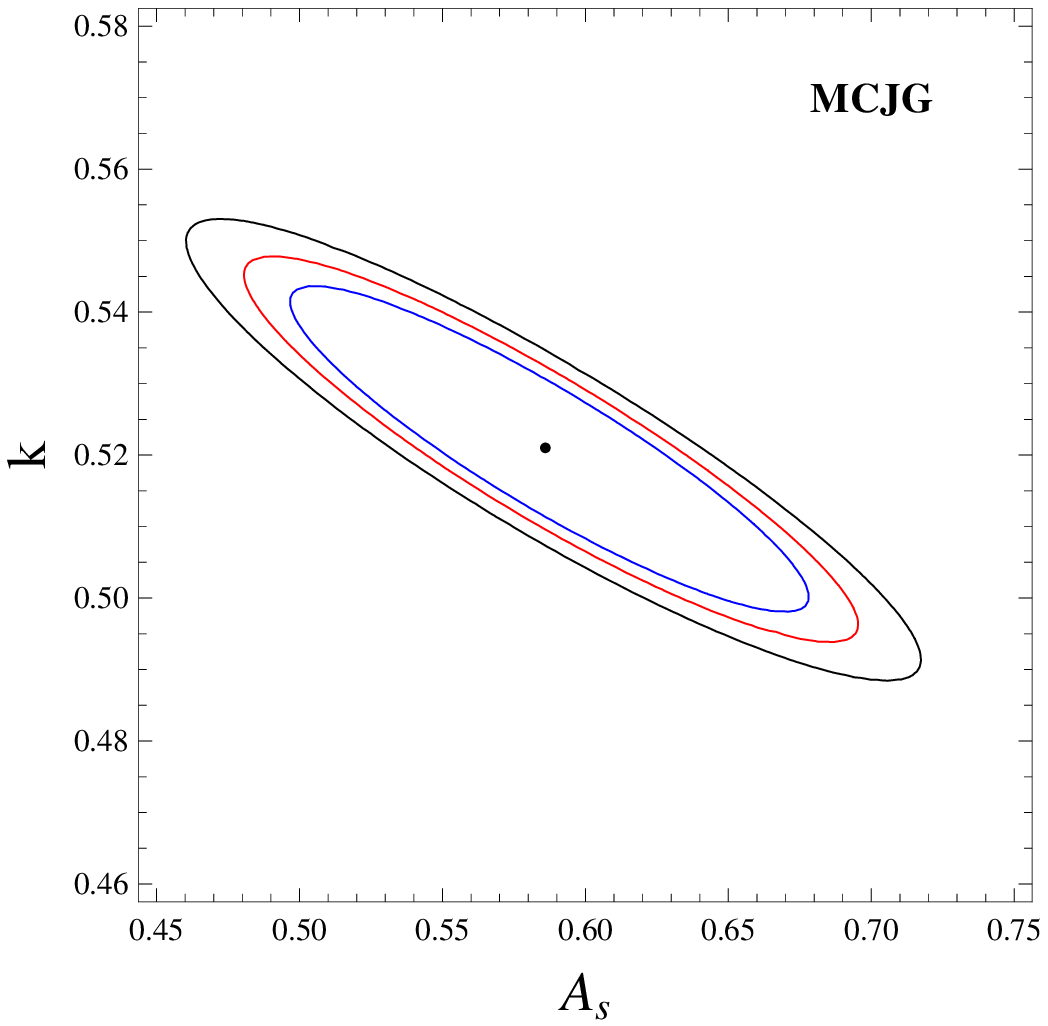}~~\\
\vspace{2mm}
~~~~~~~~~~~~~~~~~~~~~~~~~~~~~~~~~~~~Fig.3~~~~~~~~~~~~~~~~~~~~~~~~~~~~~~~~~~~~~~~~~~~~~~~~~~~~~~~~Fig.4
\\
\vspace{2mm} {\bf Figs.1 - 4:} {\it Contour plots of $h$ vs
$\Omega_{m0}$, $\alpha$ vs $A_{s}$, $A$ vs $A_{s}$ and $k$ vs
$A_{s}$ for MCJG by OHD+CMB+BAO+SNIa joint data analysis for
different confidence levels $1\sigma$ (blue), $2\sigma$ (red) and
$3\sigma$ (black).} \vspace{1cm}
\end{figure}

\begin{figure}
~~~~~~~~~~
\includegraphics[height=2.5in, width=2.7in]{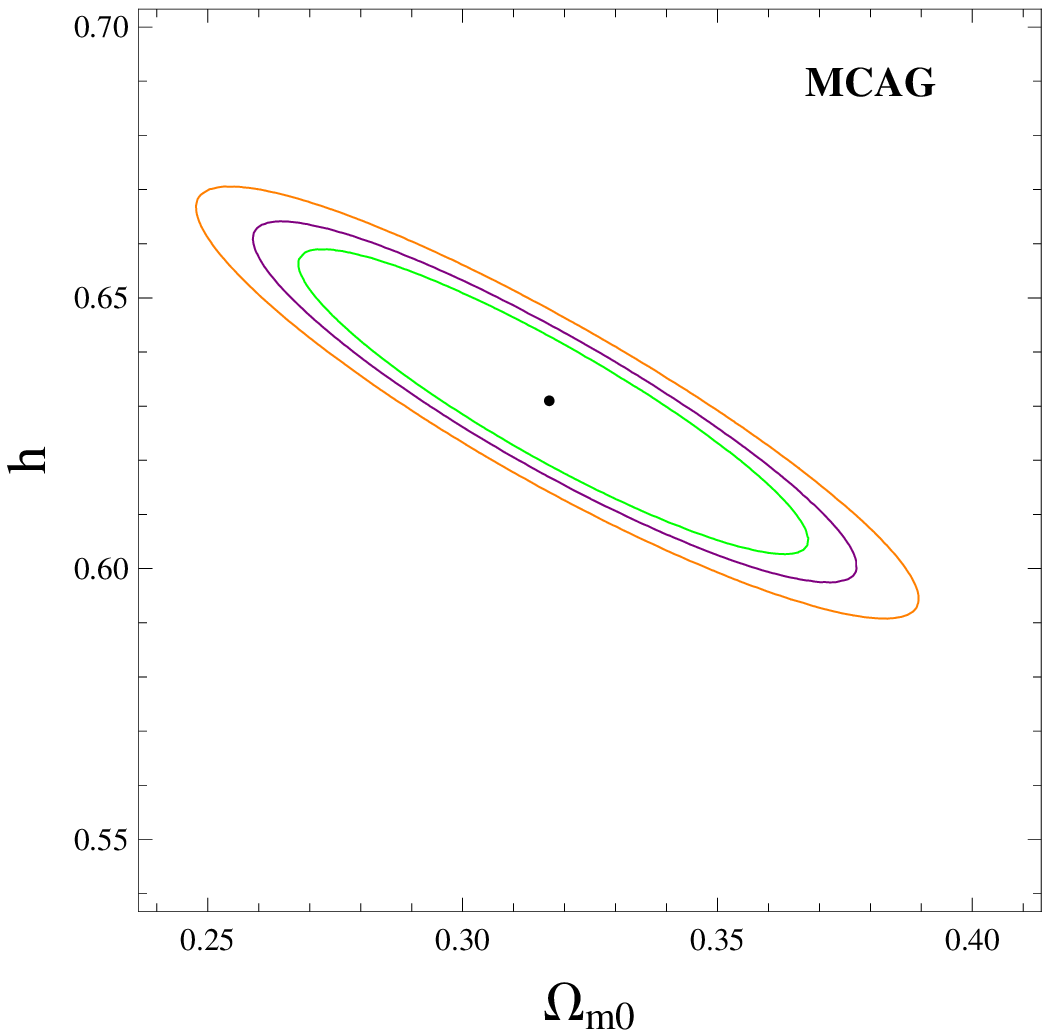}~~~~~~~
\includegraphics[height=2.5in, width=2.7in]{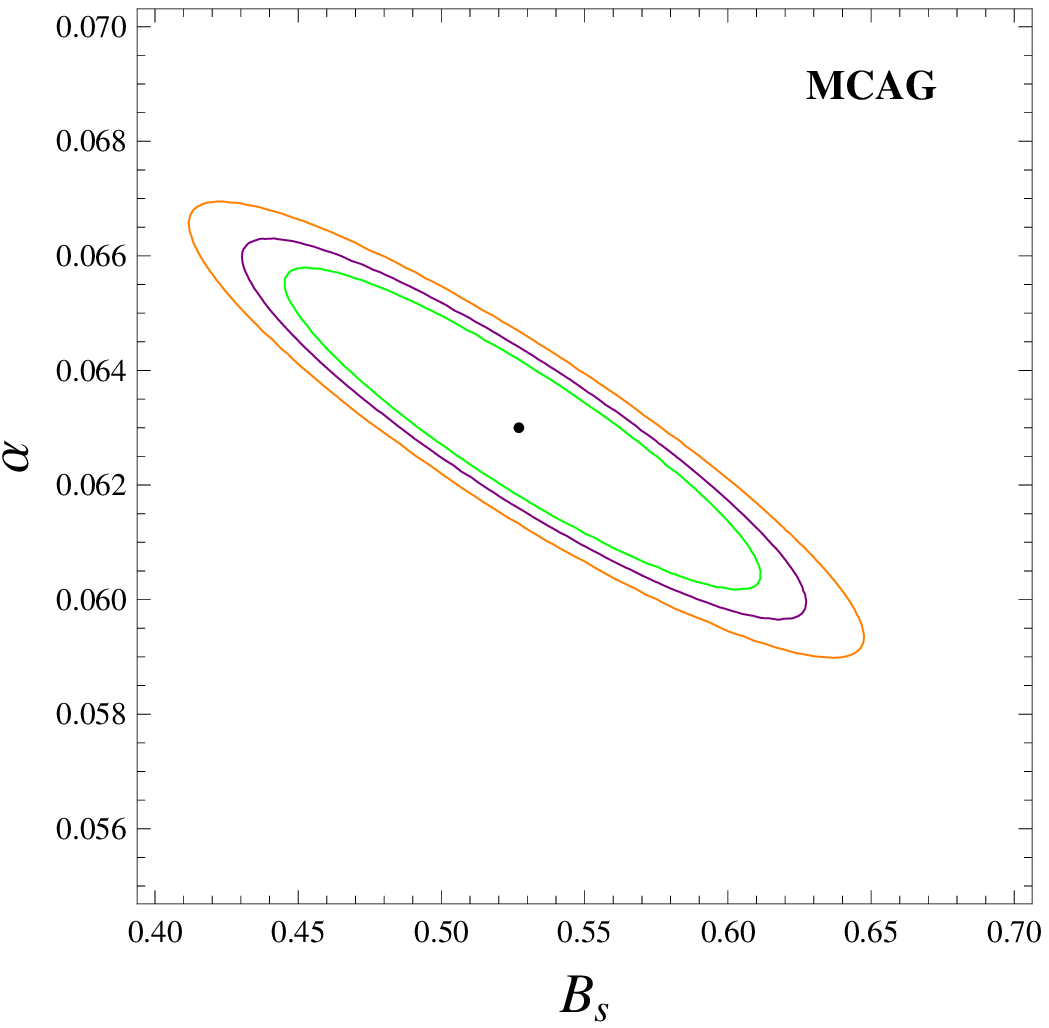}~~\\
\vspace{2mm}
~~~~~~~~~~~~~~~~~~~~~~~~~~~~~~~~~~~~Fig.5~~~~~~~~~~~~~~~~~~~~~~~~~~~~~~~~~~~~~~~~~~~~~~~~~~~~~~~~Fig.6
\\
\vspace{2mm} ~~~~~~~~~~
\includegraphics[height=2.5in, width=2.7in]{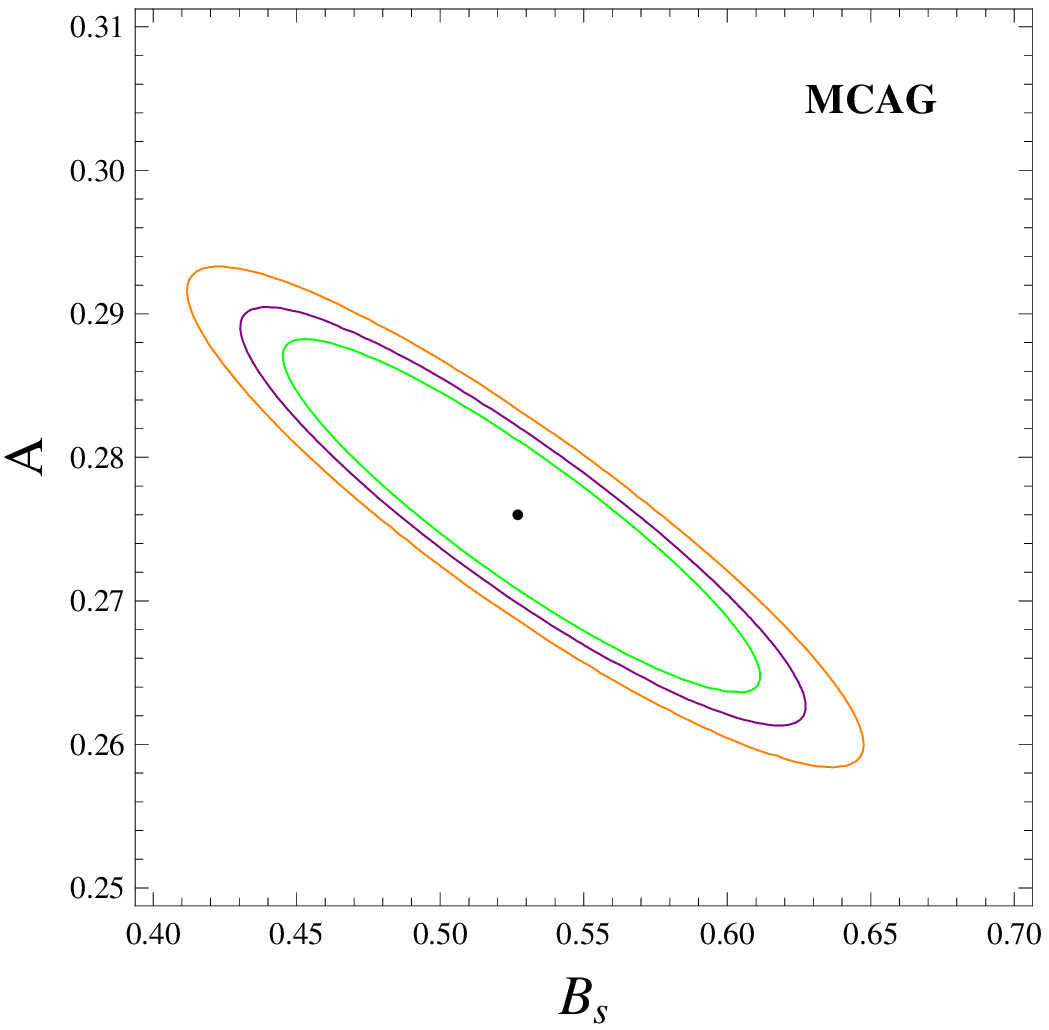}~~~~~~~
\includegraphics[height=2.5in, width=2.7in]{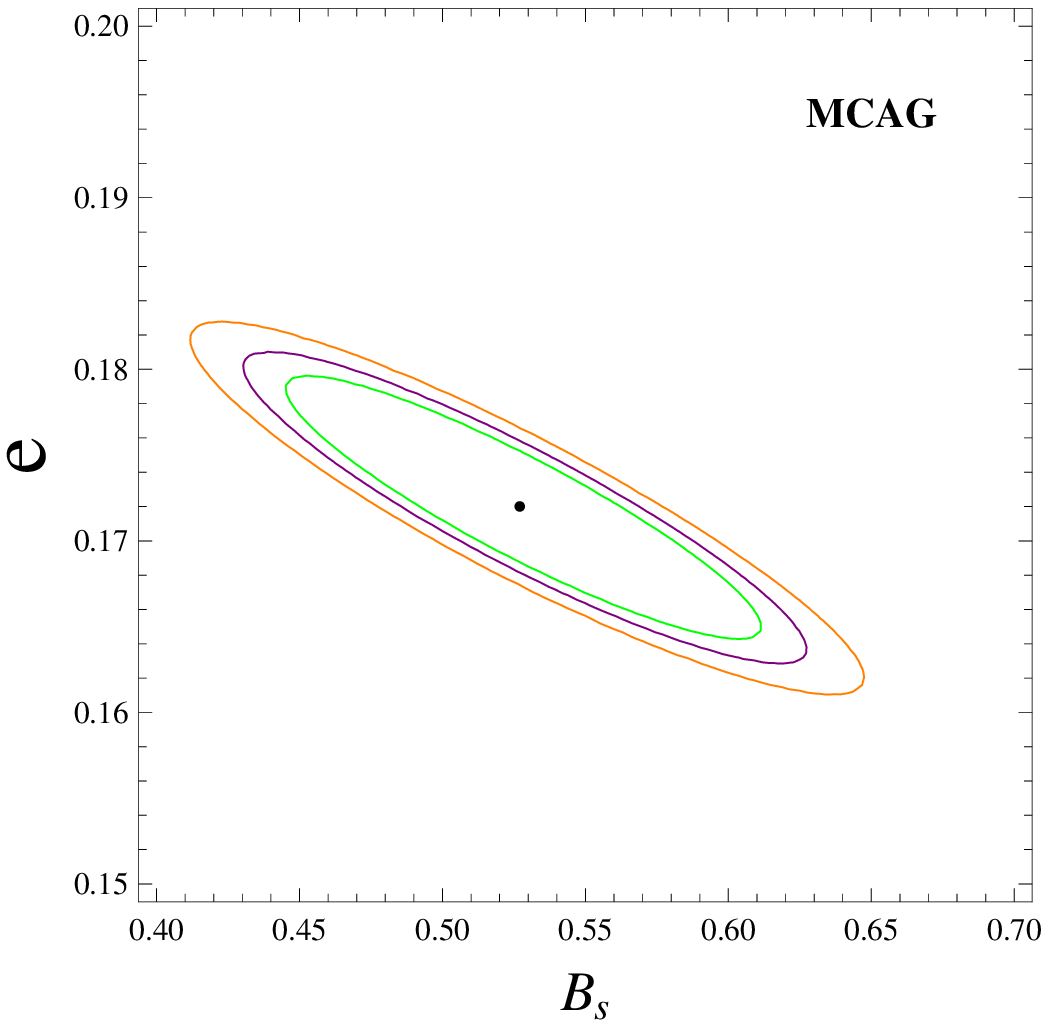}~~\\
\vspace{2mm}
~~~~~~~~~~~~~~~~~~~~~~~~~~~~~~~~~~~~Fig.7~~~~~~~~~~~~~~~~~~~~~~~~~~~~~~~~~~~~~~~~~~~~~~~~~~~~~~~~Fig.8
\\
\vspace{2mm}~~~~~~~~~~~~~~~~~~~~~~~~~~~~~~~~~~~~~~~
\includegraphics[height=2.5in, width=2.7in]{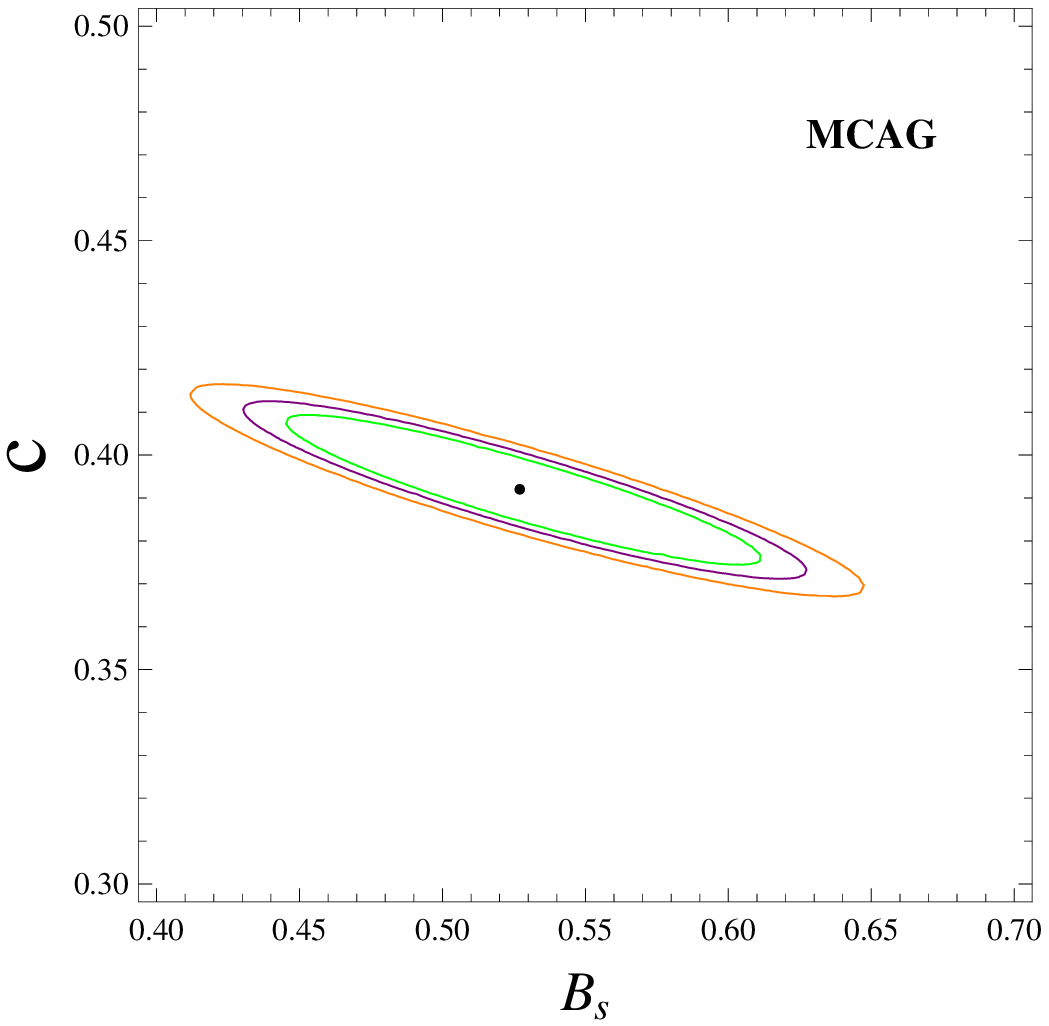}~~\\
\vspace{2mm}
~~~~~~~~~~~~~~~~~~~~~~~~~~~~~~~~~~~~~~~~~~~~~~~~~~~~~~~~~~~~~~~~~Fig.9~~~~~~
\\
\vspace{2mm}

{\bf Figs.5 - 9:} {\it Contour plots of $h$ vs $\Omega_{m0}$,
$\alpha$ vs $B_{s}$, $A$ vs $B_{s}$, $e$ vs $B_{s}$ and $c$ vs
$B_{s}$ for MCAG by OHD+CMB+BAO+SNIa joint data analysis for
different confidence levels $1\sigma$ (green), $2\sigma$ (purple)
and $3\sigma$ (orange).} \vspace{1cm}
\end{figure}

\subsection{Results}

In the MCJG and MCAG models, the best fit values of the parameters
can be obtained by $\chi^2$ minimum test using the maximum
likelihood estimation. The parameter of the radiation density is
given by the formula $\Omega_{r0}=\Omega_{m0}/(1+z_{eq})$ where
$z_{eq}=2.5\times 10^{4}\Omega_{m0}h^{2}(T_{cmb}/2.7K)^{-4}$. Now
we choose $\Omega_{rad 0}=8.14 \times 10^{-5}$. Using
OHD+CMB+BAO+SNIa joint data analysis, the best fit parameters for
MCJG and MCAG modes are obtained by (i) $\Omega_{m0}=0.328$,
$h=0.657$, $\alpha=0.066$, $A=0.297$, $A_{s}=0.586$, $k=0.521$,
$\chi^{2}_{min}=657.381$ and (ii) $\Omega_{m0}=0.317$, $h=0.631$,
$\alpha=0.063$, $A=0.276$, $B_{s}=0.527$, $e=0.172$, $c=0.392$,
$\chi^{2}_{min}=632.197$ respectively. The best fit results with
errors of the model parameters are also given in Tables 1 and 2.
In the figures 1 - 4, we have drawn the contour plots of $h$ vs
$\Omega_{m0}$, $\alpha$ vs $A_{s}$, $A$ vs $A_{s}$ and $k$ vs
$A_{s}$ for MCJG by OHD+CMB+BAO+SNIa joint data analysis for
$1\sigma$, $2\sigma$ and $3\sigma$ confidence levels. Also in the
figures 5 - 9, we have plotted the contours of $h$ vs
$\Omega_{m0}$, $\alpha$ vs $B_{s}$, $A$ vs $B_{s}$, $e$ vs $B_{s}$
and $c$ vs $B_{s}$ for MCAG model. Since $\Lambda$CDM model is the
reference model, so the values of $\triangle$AIC, $\triangle$BIC
and $\triangle$DIC for MCJG and MCAG models can be measured
relative to the $\Lambda$CDM model. For $\Lambda$CDM model, we
know that $\triangle$AIC=$\triangle$BIC=$\triangle$DIC=0
\cite{XZ}. But for MCJG and MCAG models, we found (i)
$\triangle$AIC=1.293, $\triangle$BIC=5.824, $\triangle$DIC=1.096
and (ii) $\triangle$AIC=1.131, $\triangle$BIC=5.795,
$\triangle$DIC=1.072 which lie on the favorable ranges. So we can
conclude that the
MCJG and MCAG models are observationally viable models.\\

\[
\begin{tabular}{|c|c|c|c|c|c|c|c|}
\hline
Model &$\Omega_{m0}$& $h$  & $\alpha$ & $A$ & $A_{s}$ & $k$ & $\chi^{2}_{min}$\\

\hline
& & & & & & &\\
  MCJG      & $0.328^{+0.049}_{-0.047}$     & $0.657^{+0.028}_{-0.027}$        &  $0.066^{+0.0025}_{-0.0026}$      &  $0.297^{+0.011}_{-0.012}$     & $0.586^{+0.09}_{-0.08}$   &    $0.521^{+0.019}_{-0.021}$  &  $657.381$  \\
& & & & & & &\\

\hline

\end{tabular}
\]
{\bf Table 1:} {\it Best fit results of the MCJG model parameters
by OHD+CMB+BAO+SNIa joint data analysis. \vspace{1mm}}

\[
\begin{tabular}{|c|c|c|c|c|c|c|c|c|}
\hline
Model~ & $\Omega_{m0}$ & $h$  & $\alpha$ & $A$ & $B_{s}$ & $e$ & $c$ & $\chi^{2}_{min}$\\

\hline
& & & & & & & &\\
  MCAG~      & $0.317^{+0.043}_{-0.049}$     & $0.631^{+0.024}_{-0.023}$        &  $0.063^{+0.0025}_{-0.0023}$      &  $0.276^{+0.012}_{-0.013}$     & $0.527^{+0.075}_{-0.077}$   &    $0.172^{+0.007}_{-0.008}$  & $0.392^{+0.018}_{-0.017}$ & $632.197$\\
& & & & & & & &\\

\hline

\end{tabular}
\]
{\bf Table 2:} {\it Best fit results of the MCAG model parameters
by OHD+CMB+BAO+SNIa joint data analysis. \vspace{1mm}}

\section{Evolutions of Cosmological and Cosmographical Parameters}

In this section, we'll study the nature of the cosmological and
cosmographical parameters due to taking the best fit values of the
parameters of MCJG and MCAG models. The propagations of MCJG and
MCAG for the evolution of the universe can be determined by the
equation of state parameter $W$. From figure 10, we see that $W$
decreases from positive level $W>0$ to $W\sim 0$, i.e., radiation
to dust models can be generated for both MCJG and MCAG. Then $W$
decreases and keeps negative sign ($-1<W<0$), i.e., quintessence
models can be generated. Here $W<-1/3$ shows the dark energy phase
of the universe. Finally, $W$ reaches to $-1$ as $z\sim -1$, i.e.,
the universe goes to the $\Lambda$CDM stages due to the
contributions of MCJG and MCAG. At the early stage of the
universe, the value of $W$ for MCAG is higher than MCJG.
Obviously, at $z=0$, the value of $W$ lies within $-1$ and $-0.8$.
So at present, our universe is accelerating, which may be caused
by both MCJG and MCAG. Since $W\nless -1$, so phantom crossing
cannot occur in our considered MCJG and MCAG models.

In order to deceleration or acceleration phase of the universe, we
need to study the deceleration parameter, which is given as
$q=-\frac{\ddot{a}}{aH^{2}}$. For the deceleration phase of the
universe, $q>0$ and for the acceleration phase of the universe,
$q<0$. The deceleration parameter has been drawn in figure 11 for
both MCJG and MCAG. As time passes, we see that $q$ decreases from
positive level to negative level, i.e., deceleration to
acceleration transition occurs to the universe. There are another
parameters named as cosmographical parameters for understanding
the past and future evolution of the universe. The cosmography is
the study of scale factor by expanding it by Taylor series with
respect to the cosmic time. The cosmographical parameter like jerk
($J$), snap ($S$), and lerk ($L$) parameters are
\cite{Pan,Riv,Mand,Pac}
\begin{eqnarray}
J=\frac{\dddot{a}}{aH^{3}}=(1+z)\frac{dq}{dz}+q(1+2q),\\
S=\frac{a^{(4)}}{aH^{4}}=-(1+z)\frac{dJ}{dz}-J(2+3q),\\
L=\frac{a^{(5)}}{aH^{5}}=-(1+z)\frac{dS}{dz}-S(3+4q)
\end{eqnarray}
So the cosmographical parameters are higher-order derivatives of
deceleration parameter $q$. We have drawn the jerk ($J$), snap
($S$) and lerk ($L$) parameters in figures 12, 13 and 14
respectively for both MCJG and MCAG models. The parameters $J$ and
$L$ both decrease sharply from some positive values to near-zero
upto $z\sim 1$ and then increase sharply upto some positive
values. The sharpness of decrease and increase curves for MCAG is
higher than MCJG. On the other hand, the parameter $S$ increases
from negative level to positive level, and the transition occurs
near $z=1$. The sharpness of decrease and increase curves for MCAG
is higher than MCJG.

Another important parameters are statefinder parameters which are
distinguishing different dark energy models. The statefinder
parameters $\{r,s\}$ are defined as \cite{Sah1,Sah2}
\begin{equation}
r=\frac{\dddot{a}}{aH^{3}}~,~s=\frac{r-1}{3(q-0.5)}
\end{equation}
For $\Lambda$CDM model, $\{r,s\}=\{0,1\}$ and for SCDM model,
$\{r,s\}=\{1,1\}$. The $\{r,s\}$ trajectory is drawn in figure 15.
We see that when $r$ increases, $s$ always decreases. Two branches
of the trajectory found in the diagram. Two branches intersect at
$r=1,~s=0$. For $s>0$, the value of $r$ is $<1$. In this region,
the lower value of $r$ is $\sim 0.4$ and the corresponding upper
value of $s$ is $\sim 0.2$. But for $s<0$, the value of $r$ is
$>1$ for both MCJG and MCAG models.

The $Om$ diagnostic of dark energy is the first derivative of the
scale factor through the Hubble parameter, was introduced to
differentiate $\Lambda$CDM from other DE models, which is defined
as \cite{varun3}
\begin{equation}
Om(x)=\frac{E^{2}(x)-1}{x^{3}-1}
\end{equation}
where $x=z+1$. For $\Lambda CDM$ model, $Om = \Omega_{0m}=$
constant. The $Om$ diagnostic against redshift $z$ is plotted in
figure 16. We see that in the region $z>0$, the $Om$ parameter
decreases from some positive values to nearly zero around $z\sim
0$. But for $z<0$, the $Om$ parameter obeys the negative signature
for both MCJG and MCAG.

Finally, we can examine whether the MCJG and MCAG are stable or
not for the best fit values of the parameters. So to study the
stability of MCJG and MCAG against small perturbation, we can
write the squared speed of sound in the form
$v_{s}^{2}=\frac{dp}{d\rho}$. If $0<v_{s}^{2}<1$, we can say that
the model is classically stable while $v_{s}^{2}<0$ or
$v_{s}^{2}>1$ represents classically unstable. The square speed of
sound $v_{s}^{2}$ is drawn in figure 17. We observe that
$v_{s}^{2}$ decreases smoothly but lies within 0.2 and 0.82. So we
can conclude that MCJG and MCAG are classically stable due to
taking the best fit values of the model's parameters.

\begin{figure}
~~~~~~~~~~
\includegraphics[width=2.7in]{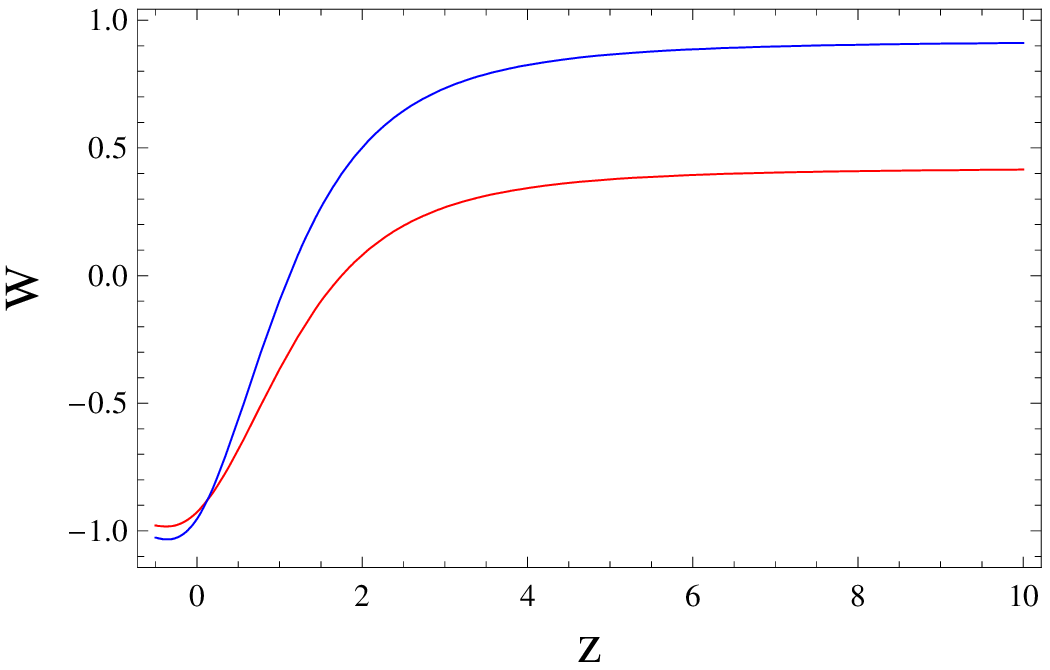}~~~~~~~
\includegraphics[width=2.7in]{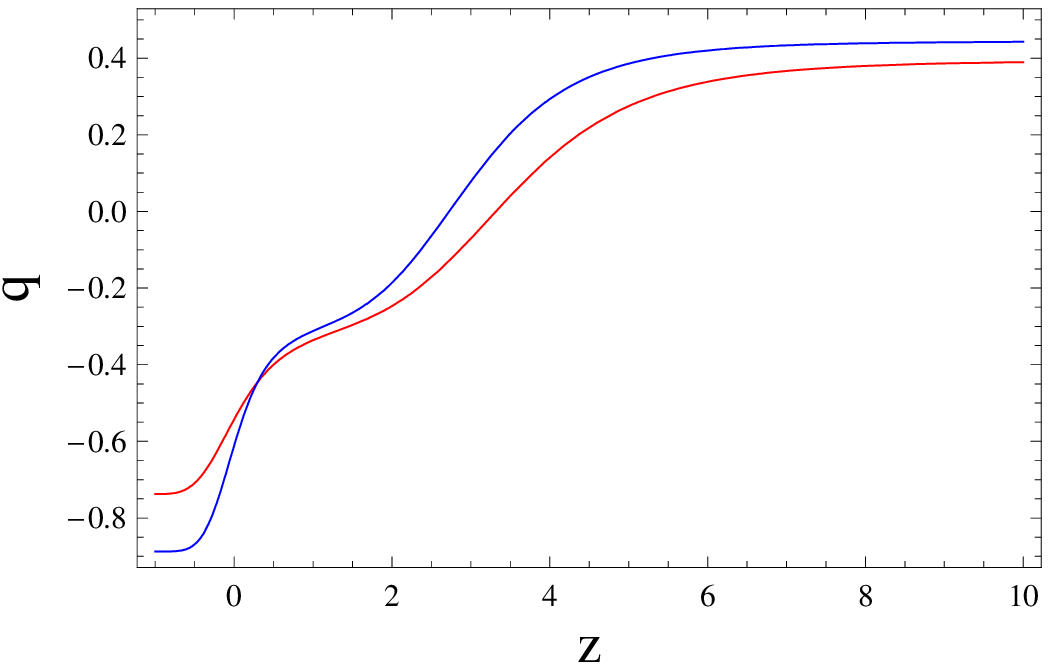}~~\\
\vspace{2mm}
~~~~~~~~~~~~~~~~~~~~~~~~~~~~~~~~~~~~Fig.10~~~~~~~~~~~~~~~~~~~~~~~~~~~~~~~~~~~~~~~~~~~~~~~~~~~~~~~~Fig.11
\\
\vspace{2mm}~~~~~~~~~~
\includegraphics[width=2.7in]{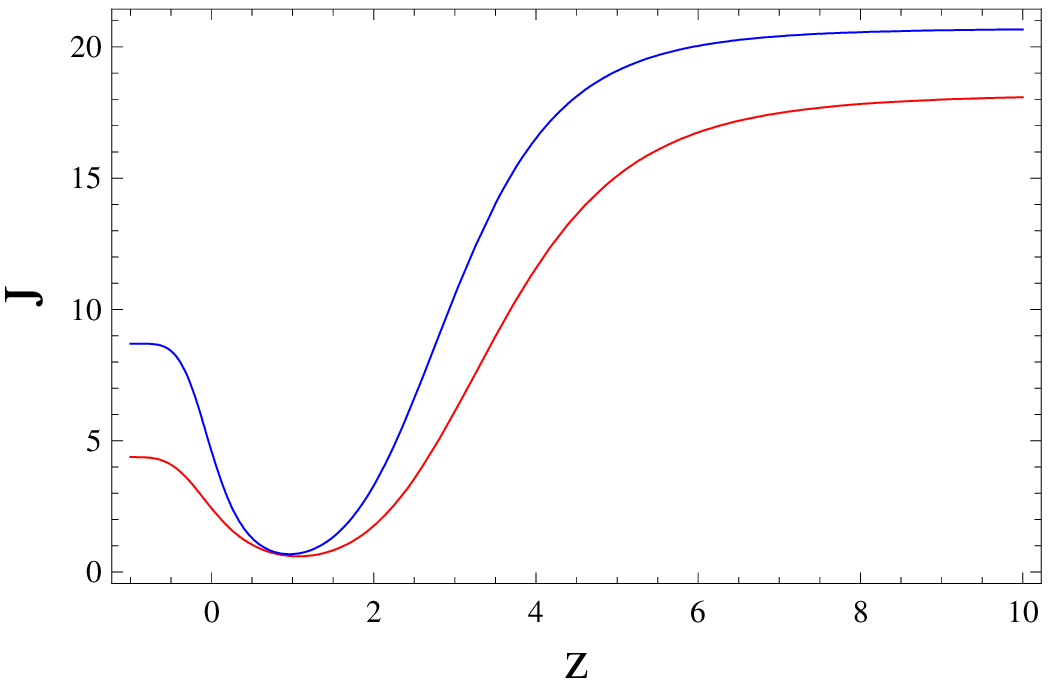}~~~~~~~
\includegraphics[width=2.7in]{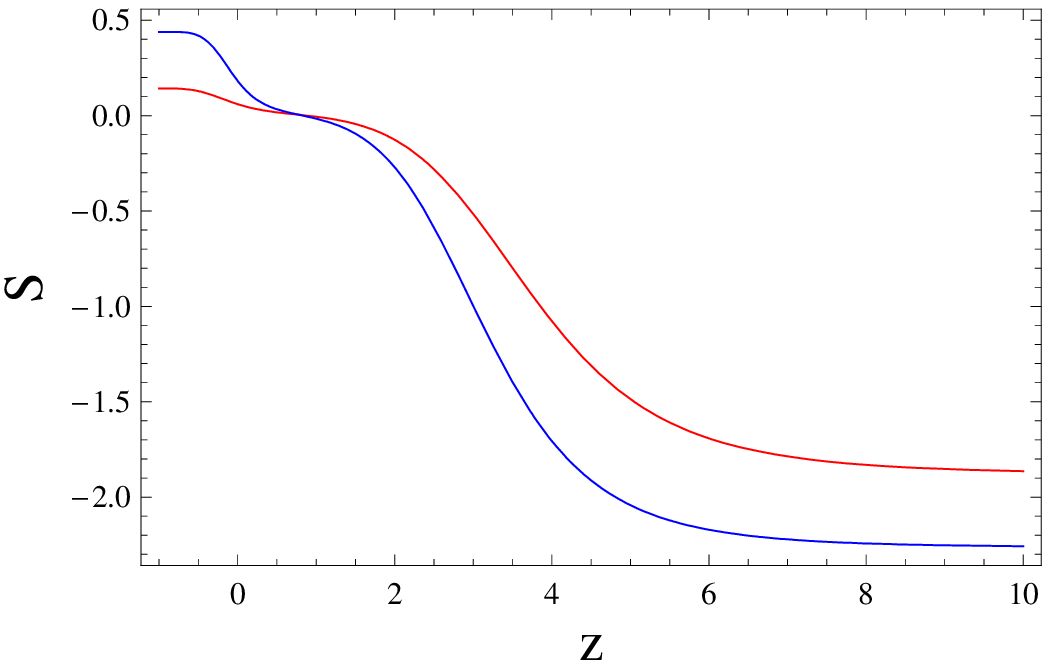}~~\\
\vspace{2mm}
~~~~~~~~~~~~~~~~~~~~~~~~~~~~~~~~~~~~Fig.12~~~~~~~~~~~~~~~~~~~~~~~~~~~~~~~~~~~~~~~~~~~~~~~~~~~~~~~~Fig.13
\\
\vspace{2mm}~~~~~~~~~~
\includegraphics[width=2.7in]{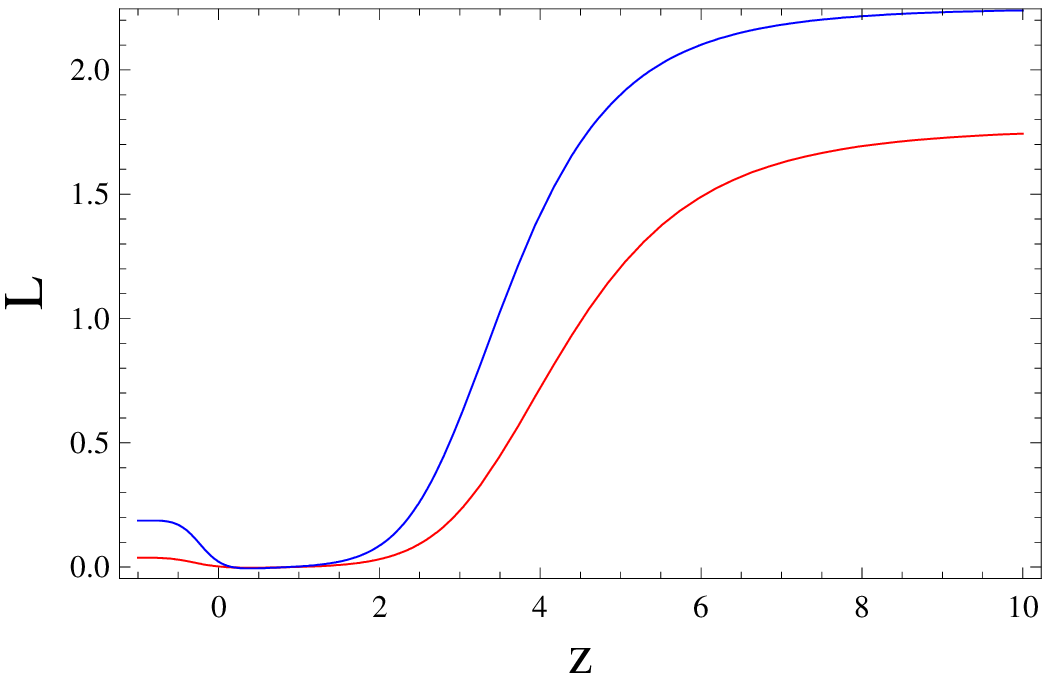}~~~~~~~
\includegraphics[width=2.7in]{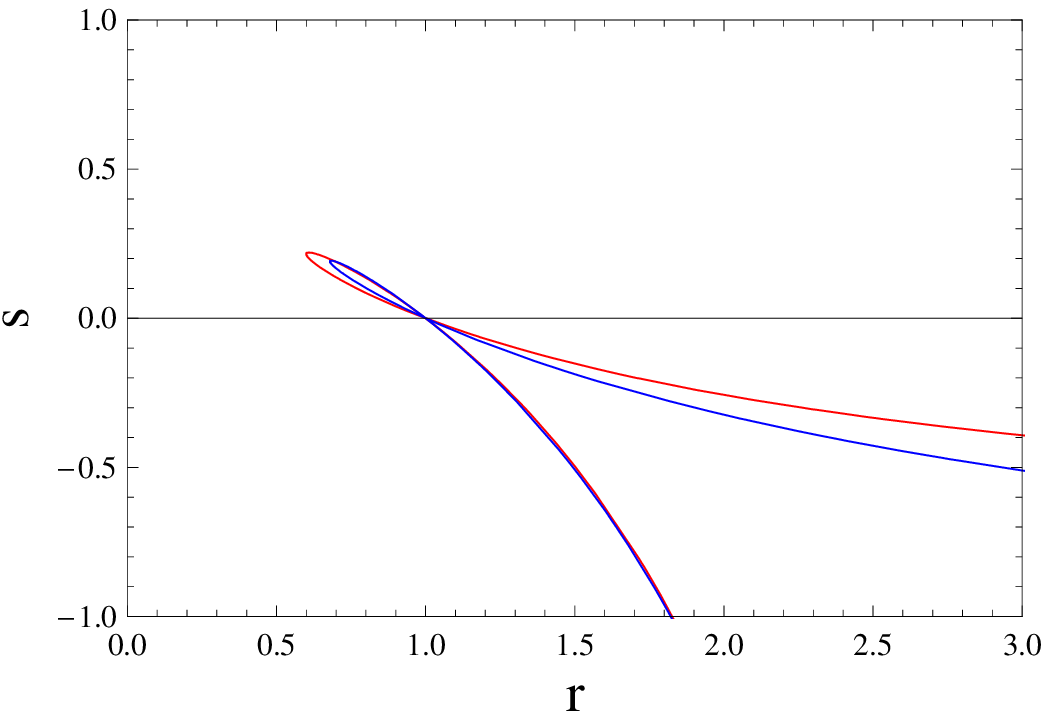}~~\\
\vspace{2mm}
~~~~~~~~~~~~~~~~~~~~~~~~~~~~~~~~~~~~Fig.14~~~~~~~~~~~~~~~~~~~~~~~~~~~~~~~~~~~~~~~~~~~~~~~~~~~~~~~~Fig.15
\\
\vspace{2mm}~~~~~~~~~~
\includegraphics[width=2.7in]{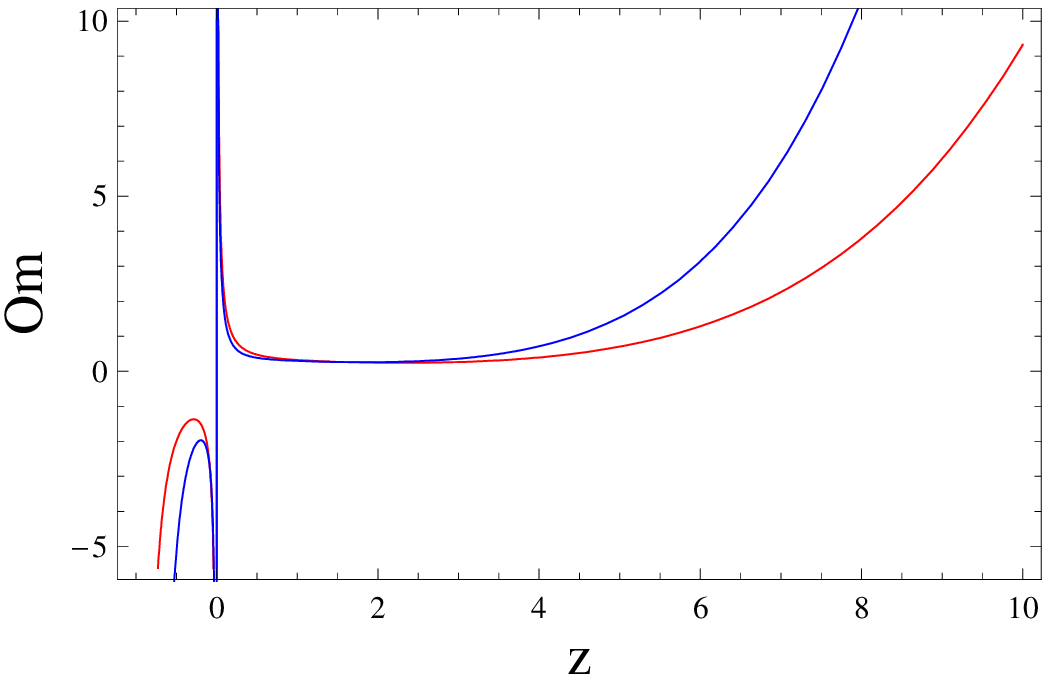}~~~~~~~
\includegraphics[width=2.7in]{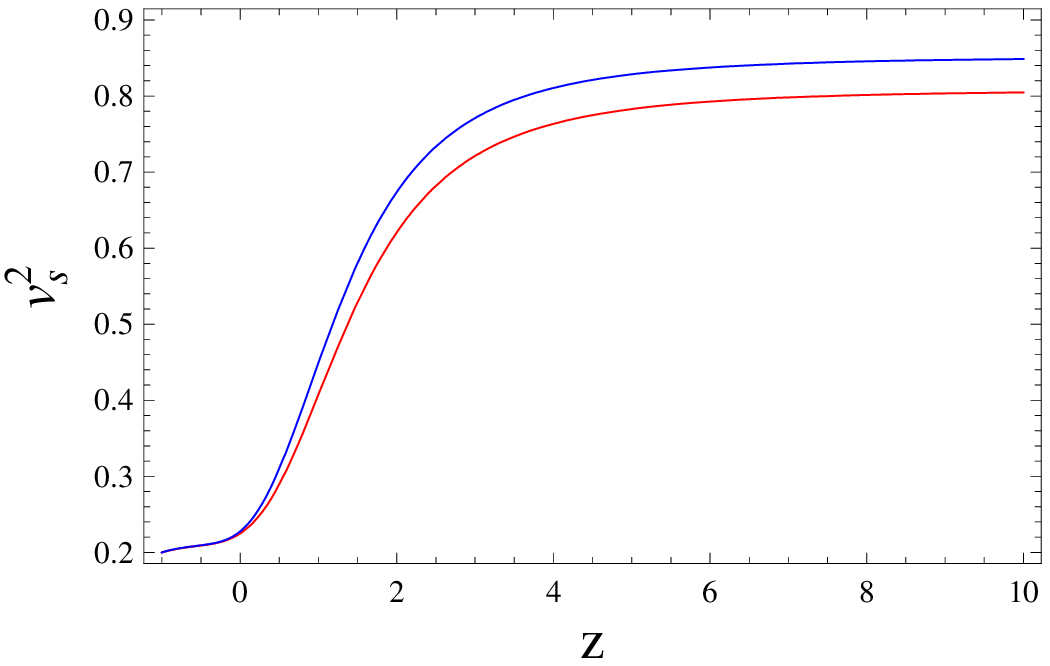}~~\\
\vspace{2mm}
~~~~~~~~~~~~~~~~~~~~~~~~~~~~~~~~~~~~Fig.16~~~~~~~~~~~~~~~~~~~~~~~~~~~~~~~~~~~~~~~~~~~~~~~~~~~~~~~~Fig.17
\\
\vspace{2mm}

{\bf Figs.10 - 17:} {\it Plots of $W$, $q$, $J$, $S$, $L$ vs $z$,
$s$ vs $r$, $Om$ and $v_{s}^{2}$ vs $z$ for MCJG (red line) and
MCAG (blue line).} \vspace{1cm}
\end{figure}

\section{\normalsize\bf{Discussions and Conclusions}}

We have considered the flat FRW model of the universe and reviewed
the modified Chaplygin gas as the fluid source. Associated with
the scalar field model, we have determined the Hubble parameter as
a generating function in terms of the scalar field. Instead of
hyperbolic function, we have taken Jacobi elliptic function and
Abel function in the generating function and obtained modified
Chaplygin-Jacobi gas (MCJG) and modified Chaplygin-Abel gas (MCAG)
equation of states, respectively. Next, we have assumed that the
universe filled in dark matter, radiation, and dark energy. The
sources of dark energy candidates are assumed as MCJG and MCAG. We
have constrained the model parameters by recent observational data
analysis. Using $\chi^{2}$ minimum test (maximum likelihood
estimation), we have determined the best fit values of the model
parameters by OHD+CMB+BAO+SNIa joint data analysis. In the figures
1 - 4, we have drawn the contour plots of $h$ vs $\Omega_{m0}$,
$\alpha$ vs $A_{s}$, $A$ vs $A_{s}$ and $k$ vs $A_{s}$ for MCJG
model in $1\sigma$, $2\sigma$ and $3\sigma$ confidence levels.
Also, in figures 5 - 9, we have plotted the contours of $h$ vs
$\Omega_{m0}$, $\alpha$ vs $B_{s}$, $A$ vs $B_{s}$, $e$ vs $B_{s}$
and $c$ vs $B_{s}$ for MCAG model. To examine the viability of the
MCJG and MCAG models, we have determined the values of the
deviations of information criteria like $\triangle$AIC,
$\triangle$BIC and $\triangle$DIC. Since $\Lambda$CDM model is the
reference model, so the values of $\triangle$AIC, $\triangle$BIC,
and $\triangle$DIC for MCJG and MCAG models can be measured
relative to the $\Lambda$CDM model. Our computational values of
$\triangle$AIC, $\triangle$BIC, and $\triangle$DIC lie on the
favorable ranges. So our constructed MCJG and MCAG are
observationally favorable models.\\

The evolutions of cosmological and cosmographical parameters (like
equation of state, deceleration, jerk, snap, lerk, statefinder,
$Om$ diagnostic) have been studied using best-fit values of MCJG
and MCAG parameters. From figure 10, we have seen that the EoS
parameter $W$ decreases from positive level $W>0$ to $W=-1$, i.e.,
radiation to $\Lambda$CDM models can be generated for both MCJG
and MCAG. But at the early stage of the universe, the value of $W$
for MCAG is higher than MCJG. At present, our universe is
accelerating, which may be caused by MCJG and MCAG. Since $W\nless
-1$, so phantom divide cannot occur in our considered MCJG and
MCAG models. Also, for both MCJG and MCAG models, from figure 11,
we have seen that the deceleration parameter $q$ decreases from
positive level to negative level, i.e., deceleration to
acceleration phase transition occurs to the universe. We have
drawn the jerk ($J$), snap ($S$), and lerk ($L$) parameters in
figures 12, 13, and 14, respectively, for both MCJG and MCAG
models. The parameters $J$ and $L$ decrease sharply from some
positive values to near-zero upto $z\sim 1$ and then increase
sharply upto some positive values. The sharpness of decrease and
increase curves for MCAG is higher than MCJG. On the other hand,
$S$ increases from negative level to positive level,  and the
transition occurs near $z=1$. The sharpness of decrease and
increase of the trajectories for MCAG is higher than MCJG. From
figure 15, we have seen that when $r$ increases, $s$ always
decreases. Two branches of the trajectory found in the $(r,s)$
diagram. Two branches intersect at $r=1,~s=0$. For $s>0$, the
value of $r$ is $<1$. In this region, the lower value of $r$ is
$\sim 0.4$ and the corresponding upper value of $s$ is $\sim 0.2$.
But for $s<0$, the value of $r$ is $>1$ for both MCJG and MCAG
models. On the other hand, from figure 16, we have seen that in
the region $z>0$, the $Om$ parameter decreases from some positive
values to nearly zero around $z\sim 0$, while for $z<0$, $Om$
parameter obeys negative signature for both MCJG and MCAG. To
check the classical stability of the models, we have examined the
values of square speed of sound $v_{s}^{2}$ in the interval
$(0,1)$ for expansion of the universe. From figure 17, we have
observed that $v_{s}^{2}$ decreases smoothly but lies within 0.2
and 0.82. So we can conclude that MCJG and MCAG are both classically stable. \\

\end{document}